\documentclass[twocolumn,trackchanges,times,tighten]{aastex63}

\usepackage[utf8]{inputenc}
\usepackage{amsmath}
\usepackage{multirow} 
\usepackage[ruled,vlined]{algorithm2e}
\usepackage{array}
\usepackage{color}

\usepackage{mathrsfs}
\usepackage{xcolor}
\definecolor{light-gray}{gray}{0.95}
\newcommand{\code}[1]{\colorbox{light-gray}{\texttt{#1}}}

\usepackage{textcomp}
\usepackage{gensymb}
\usepackage{graphicx}

\shorttitle{Bits missing: Finding exotic pulsars using bfloat16 on NVIDIA GPUs}
\shortauthors{White J., Ad\'amek K., Roy J., Dimoudi S., Ransom S., Armour W.}

\begin{document}
\title{Bits missing: Finding exotic pulsars using bfloat16 on NVIDIA GPUs}

\correspondingauthor{Wesley Armour}
\email{wes.armour@oerc.ox.ac.uk}

\author[0000-0003-2690-6858]{Jack White}
\affiliation{Oxford e-Research Centre, Department of Engineering Science, University of Oxford, 7 Keble Road, Oxford OX1 3QG, United Kingdom}

\author[0000-0003-2797-0595]{Karel Ad\'{a}mek}
\affiliation{Oxford e-Research Centre, Department of Engineering Science, University of Oxford, 7 Keble Road, Oxford OX1 3QG, United Kingdom}
\affiliation{Department of Physics, Silesian University in Opava, Bezru\v{c}ovo n\'{a}m., 746 01, Opava, Czech Republic}

\author[0000-0002-2892-8025]{Jayanta Roy}
\affiliation{National Centre for Radio Astrophysics (NCRA-TIFR), Pune 411 007, India}

\author[0000-0002-0967-1332]{Sofia Dimoudi}
\affiliation{Centre for Advanced Instrumentation, Durham University, United Kingdom}

\author[0000-0001-5799-9714]{Scott M. Ransom}
\affiliation{National Radio Astronomy Observatory, Charlottesville, VA 22903, United States }

\author[0000-0003-1756-3064]{Wesley Armour}
\affiliation{Oxford e-Research Centre, Department of Engineering Science, University of Oxford, 7 Keble Road, Oxford OX1 3QG, United Kingdom}

\begin{abstract}

The Fourier Domain Acceleration Search (FDAS) is an effective technique for detecting faint binary pulsars in large radio astronomy datasets. This paper quantifies the sensitivity impact of reducing numerical precision in the GPU accelerated FDAS pipeline of the AstroAccelerate software package. The prior implementation used IEEE-754 single-precision in the entire binary pulsar detection pipeline, spending a large fraction of the runtime computing GPU accelerated FFTs. AstroAccelerate has been modified to use bfloat16 (and IEEE-754 double-precision to provide a ``gold standard" comparison) within the Fourier domain convolution section of the FDAS routine. Approximately 20,000 synthetic pulsar filterbank files representing binary pulsars were generated using SIGPROC with a range of physical parameters. They have been processed using bfloat16, single and double-precision convolutions. All bfloat16 peaks are within 3\% of the predicted signal-to-noise ratio of their corresponding single-precision peaks. Of 14,971 ``bright" single-precision fundamental peaks above a power of 44.982 (our experimentally measured highest noise value), 14,602 (97.53\%) have a peak in the same acceleration and frequency bin in the bfloat16 output plane, whilst in the remaining 369 the nearest peak is located in the adjacent acceleration bin. There is no bin drift measured between the single and double-precision results. The bfloat16 version of FDAS achieves a speedup of approximately 1.6x compared to single-precision. A comparison between AstroAccelerate and the PRESTO software package is presented using observations collected with the GMRT of PSR J1544+4937, a 2.16ms black widow pulsar in a 2.8 hour compact orbit.

\end{abstract}
\keywords{Pulsar --- FFT --- Astronomy data reduction --- Computational astronomy}

\section{Introduction}
\label{sec:introduction}

Binary pulsar systems provide the means to test general relativity via observations of their emissions in the radio spectrum. Described by  \cite{1982ApJ...253..908T}, these systems include at least one pulsar, in orbit with another compact object. 
As the pulsar accelerates towards and away from the observer, the Doppler shift modulates the otherwise periodic pulsar signal (in its own reference frame) into a pulse train with varying period in the frame of the observer, the nature of which depends on the shape of the orbit. 
This effect reduces the sensitivity of the Fourier transform to the pulse trains emitted by these objects as it spreads the power of the signal across many frequencies, rendering the signals indistinguishable from the background noise.

Due to the accurate clock-like property of the pulsar embedded in the binary system, binary pulsars (particularly millisecond pulsars) are of interest to astronomers globally. They are good references to trace the evolution of the binary system \cite{1991PhR...203....1B} and they are useful for testing the theory of gravity \cite{GravityTest}. Thus significant efforts are being made in searching for such accelerated binary systems using time-domain radio surveys. Some recent examples of relevant research include \cite{10.1093/mnras/stab790}, \cite{2021}, \cite{Kansabanik_2021}, \cite{2022MNRAS.510.1393M} and \cite{2021arXiv210308410R}.

When looking to next generation radio telescopes, time-domain data rates will be of the order of TB/s. For example, SKA is expected to produce ~800GB/s \cite{levin2017pulsar}. Due to the volume of data, it is not feasible to store and so analysis must be performed in real-time, as data is captured. This motivates the investigation of using reduced precision computations in computationally demanding parts of the pulsar detection pipeline.

Within such detection pipelines, search techniques for binary pulsar systems that aim to capture all of the power emitted by the pulsar and collect it into a single detection, are typically the most computationally demanding.

Two different methods are favoured for searching for binary systems. The first is via time domain methods  \cite{1984ApJ...279..157M}, where the observed time series from a radio telescope is repeatedly Doppler shifted and Fourier transformed in an attempt to move the observed data into a frame whereby the emitter appears to the observer to be stationary.

The second method relies on convolving the frequency spectrum of the observed time series with templates that represent the frequency domain signatures of pulsars with given Doppler shifts. The computation of this approach can be accelerated via the convolution theorem and GPU accelerated Fast Fourier Transforms (FFTs).

FDAS (Fourier Domain Acceleration Search) is an implementation of the latter approach, a search technique for detecting Doppler-shifted binary pulsar signals in time series datasets, generated by radio telescopes. It was proposed in \cite{Ransom2002:FDAS} and then extended by \cite{Andersen2018:JERK} to account for nonlinear acceleration, known as Jerk Searching. AstroAccelerate implements GPU accelerated FDAS and Jerk Searching \cite{dimoudi2015pulsar}, \cite{adamek2017improved}, \cite{Sofia:2018:FDAS}, \cite{adamek2018gpu}, \cite{2020ASPC..527..671A}. In AstroAccelerate, the majority of the numerical computation in the binary pulsar detection pipeline is performed during FDAS.

The goal of this work is to establish whether switching the bulk of the numerical processing in the binary pulsar detection pipeline of AstroAccelerate to reduced (bfloat16) precision leads to acceptable sensitivity in the output, compared to the existing single-precision version. The motivation for this is to speed up the pipeline to allow the widest possible parameter search space in real time on next generation radio telescopes, or reduce the hardware requirements needed to perform such searches.

When performing FDAS, AstroAccelerate spends a significant fraction of the runtime performing GPU accelerated FFTs. The performance of a GPU accelerated FFT is limited by memory bandwidth on current generation hardware (NVIDIA Ampere), and so by reducing the numerical precision of the calculations from 32 bits to 16 bits, this bottleneck can be alleviated simply by halving the amount of data that needs to be transported to and from the GPU processing cores.

In this paper we present the results of testing a bfloat16 implementation of FDAS in AstroAccelerate with synthetic pulsar files using SIGPROC's \cite{2011ascl.soft07016L} \code{fake} filterbank generator, spanning a wide range of parameters and finally a confirmation of our findings using observations from the GMRT.

\section{Searching for binary pulsars}

\subsection{Detailed explanation}
\label{sec:detailedexplanation}

Fourier Domain Acceleration Searching uses convolution to compare the Fourier transformed time-domain signal with templates representing a range of acceleration values. A peak in the response will be seen when the signal and template overlap, indicating the presence of a pulsar with similar parameters to the template. The number of templates used depends on the search parameters desired by the radio astronomer. 

During pulsar searches, the radio astronomer is primarily interested in pulsars that haven't already been discovered. Therefore the range of search parameters must be wide enough to also include ``exotic" pulsars. This might mean, for example, pulsars with an extremely fast pulse period, combined with extreme acceleration, the latter suggesting a very tight orbit around a compact object. In such cases, more pulsar templates are required to span the extreme values of acceleration. The compute and memory requirements of FDAS increases linearly with the number of templates.

These templates will represent varying values of Doppler shift (which is a function of the particular orbital dynamics of the system), the span of the Doppler shift values defines the breadth of a binary pulsar search, along the axis known as frequency derivative or $\dot{f}$ which is related to the acceleration of the pulsar, and the Fourier bin drift $z$, as:

\begin{equation}
\label{fdotz}
Acceleration = \frac{\dot{f}c}{f} = \frac{zc}{fT^2}
\end{equation}

Where $f$ is the frequency of the pulsar and $T$ is the observation time.

The signal is convolved with all templates resulting in the $f-\dot{f}$ plane. Frequency $f$ and frequency derivative $\dot{f}$ therefore make up the primary axes of the pulsar search, a pulsar with a given pulse frequency and acceleration value will appear as a peak on this plane, with diminishing harmonics at integer multiples of both $f$ and $\dot{f}$. The summed height (power) of the peak and harmonics is typically used to calculate the signal-to-noise ratio (SNR) value of the detected pulsar against the statistics of the background noise.

\begin{figure}[htb]
\includegraphics[scale=0.3]{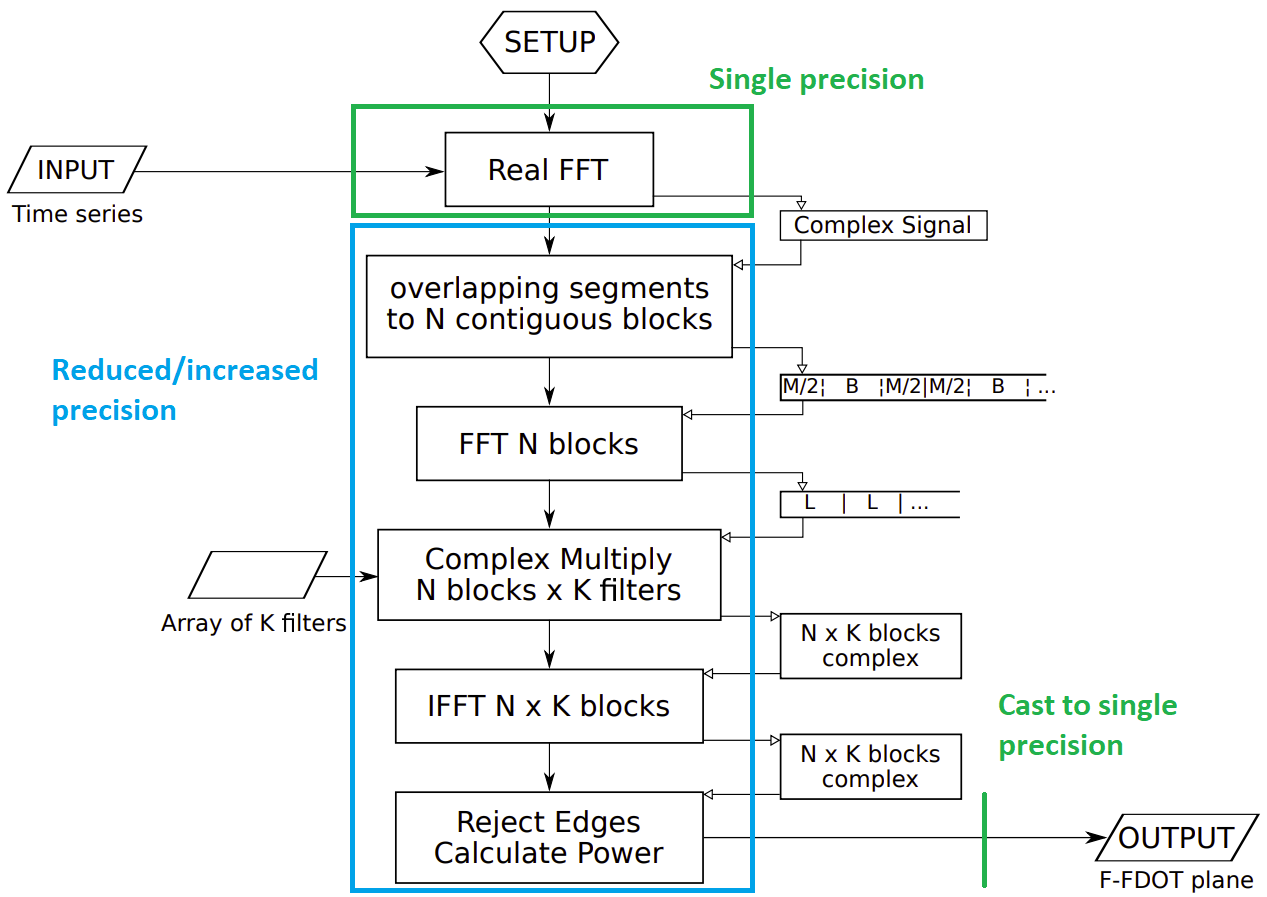}
\caption{Flowchart depicting the process of Fourier Domain matched filtering using the Overlap-Save method on a GPU. Courtesy of \cite{Sofia:2018:FDAS}}
\label{fdas_flow}
\centering
\end{figure}

The convolution of a signal and a set of templates could be completed entirely in the time domain, but a time-domain convolution involves $O(N^2)$ multiplications and $O(N^2)$ additions, leading to overall $O(N^2)$ complexity. Using the FFT and convolution theorem, this can be reduced to an overall complexity $O(NlogN)$ in the Fourier domain. The Overlap-Save method is used to perform the convolutions on the GPU, as described in \cite{Sofia:2018:FDAS}, \cite{adamek2020gpu} and Figure \ref{fdas_flow}. Briefly, this involves splitting the long input signal into windows which are separately convolved with the templates, and then the central region of each windowed convolution is saved, discarding the contaminated regions at the edge of the windows.

\subsection{FDAS vs Jerk Search}

As mentioned in Section \ref{sec:introduction}, two related methods are available when searching for binary systems in the Fourier domain, these are the Fourier Domain Acceleration Search (FDAS) and Jerk Search. FDAS makes the assumption that the acceleration of the emitting object over the observation time is approximately constant. The observation time should be no longer than approximately 10\% of the orbital period for this assumption to hold.

Jerk Searching takes into account a change in acceleration, and therefore extends the $f-\dot{f}$ plane into an $f-\dot{f}-\ddot{f}$ volume. More candidate templates are required to populate the third dimension, $\ddot{f}$, for each $\dot{f}$ value that is evaluated. Since Jerk Searching is based on the same principal of convolving templates, for this paper we have chosen to focus on the less computationally extensive FDAS to allow us to profile the numerical sensitivity of AA across a wider range of $\dot{f}$ than we would be able to in a given time with Jerk Searching.

\section{Updating the pipeline to bfloat16}

The existing version of AstroAccelerate used single-precision in all stages of FDAS (see Figure \ref{fdas_flow}). For this work we aim to profile the numerical sensitivity of the output of AstroAccelerate to changing precision in FDAS, with a focus on reducing precision to bfloat16 and comparison with IEEE-754 double-precision.

\begin{figure}[htb]
\centering
\includegraphics[scale=0.4]{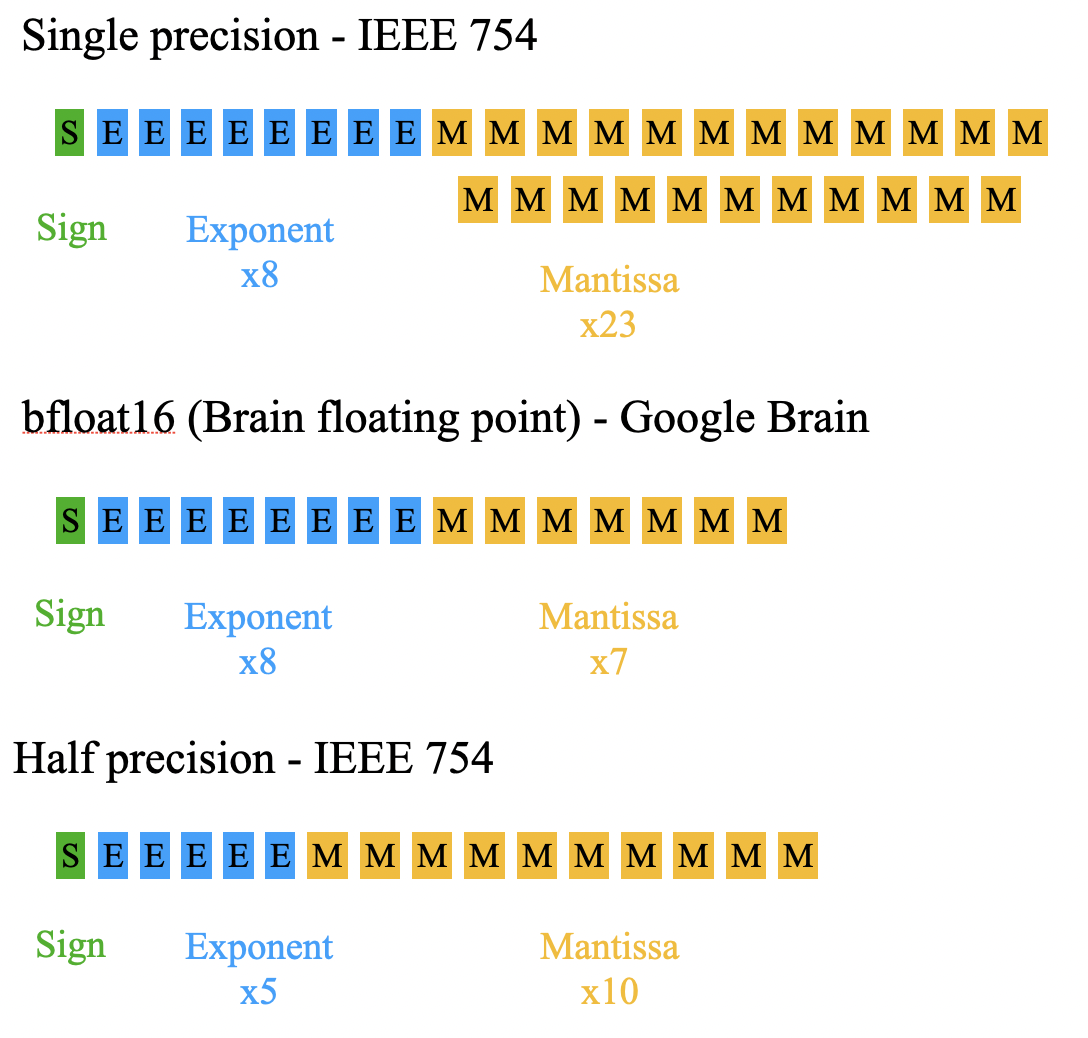}
\caption{Comparison of bit layouts of IEEE-754 single and half-precision with bfloat16}
\label{16bit_comparison}
\centering
\end{figure}

Refactoring the code to mixed precision is a straightforward process, involving replacement of the existing single-precision functions with their bfloat16 equivalents.

The `Real FFT' in Figure \ref{fdas_flow} was intentionally left in single-precision as it does not significantly affect the overall execution time (it does not scale with $K$, the number of filters). Reducing the precision of this FFT disproportionately affects the quality of the output for two reasons. Firstly, at this stage of the pipeline, the data may not span a large dynamic range, and may not be centered on zero, which highlights the weakness of bfloat16. Compared to single-precision, at large scales the spacings between bfloat16 numbers are extremely coarse, so small perturbations on the input data may not be representable without the use of single-precision. Secondly, this FFT (with a length equal to the entire input data) is sufficiently long for the numerical error, which accumulates with the larger number of calculations, to negatively impact sensitivity.

\subsection{Casting between single-precision and bfloat16}

As can be seen in Figure \ref{16bit_comparison}, in binary form a bfloat16 number can be represented in single-precision simply by appending $16\times$ zero bits to the end. Therefore single-precision and bfloat16 have approximately the same valid range ($\approx\pm3\times10^{38}$), however bfloat16 numbers are far coarser within this range.

To minimise time spent converting data once the output $f-\dot{f}$ plane had been produced, in the bfloat16 version of the code the final cast to single-precision (see Figure \ref{fdas_flow}) is completed by copying the two bytes of data for each bfloat16 number into the corresponding positions of an all zero single-precision output array.

This quick casting is only possible due to the perfect bit alignment between bfloat16 and single-precision, it is not possible with half-precision and single-precision. Quick casting all but eliminates the performance impact of implicitly casting the data using a standalone GPU kernel.

\subsection{IEEE-754 Half vs bfloat16 precision}

The valid range of any data format type is a consideration when comparing floating point formats. Half-precision (IEEE-754) has a valid range of $\pm65504$. This introduces some considerations for hypothetically using half-precision in this application.

When calculating the Fourier spectrum of a timeseries, the $0$th bin is the sum of all elements of the timeseries, often referred to as the ``DC component" of the signal. Unless the signal is perfectly centered on zero, this may mean that in long FFT calculations, the 0th bin accumulates power which takes its value outside of the $\pm65504$ bounds imposed by IEEE-754 half-precision. Depending on how the hardware handles numerical overflows, in the worst case this could introduce an \code{Inf} or \code{NaN}, while in the best case, any information outside this range would be lost. If an \code{Inf} or \code{NaN} is present in a spectrum which is subsequently inverse Fourier transformed, all values of the output will be polluted and subsequently unusable.

Given the risk of introducing overflow errors to the pipeline, we selected bfloat16 as our candidate data format for reducing the precision, which also included the benefit of quick casting.

\subsection{Template conversion}
\begin{figure}[htb]
\includegraphics[scale=0.16]{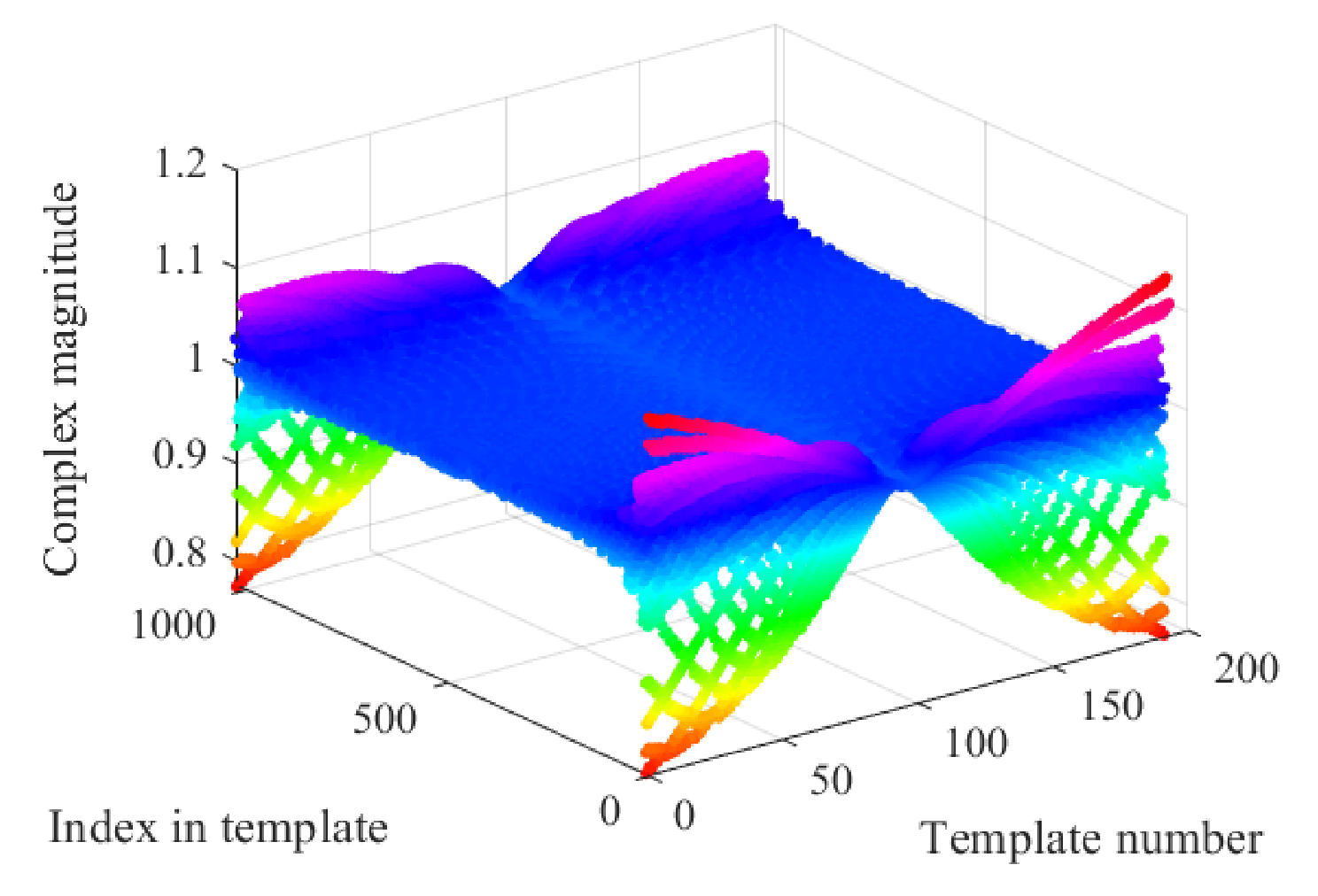}
\caption{Single-precision acceleration templates - complex magnitude plotted}
\label{single_acc_template}
\centering
\end{figure}

To convert the convolution step of FDAS to a lower precision, it is first necessary to convert the convolution templates to lower precision. These can be thought of as the candidate pulsar signatures that the code will look for in the datasets. Each template is a 1D array of complex numbers, the single-precision complex magnitudes are plotted in Figure \ref{single_acc_template}. In all experiments, the convolution templates are first generated in single-precision, and then explicitly/implicitly converted to bfloat16/double-precision by a tuned GPU kernel.

For the real component and the imaginary component, we calculate the relative difference between the single-precision and bfloat16 precision value of the template as:

\begin{equation}
\label{eqn:perc_diff}
\% = \frac{b-s}{\frac{b+s}{2}}\times100
\end{equation}

\begin{figure}[htb]
\includegraphics[scale=0.27]{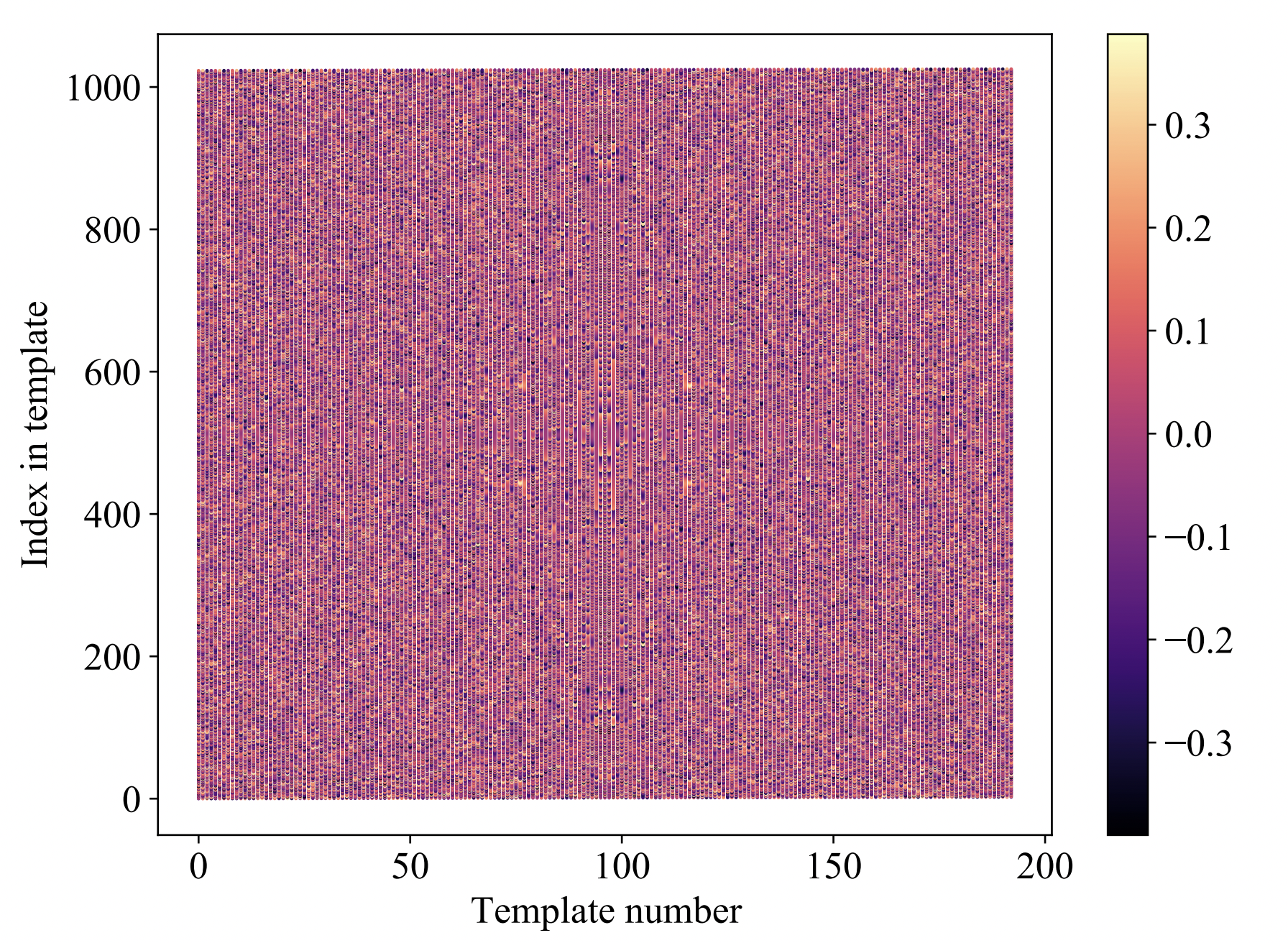}
\caption{Relative difference between real components of acceleration templates, colour scale represents relative difference as in Equation \ref{eqn:perc_diff}.}
\label{rediff_template}
\centering
\end{figure}

\begin{figure}[htb]
\includegraphics[scale=0.27]{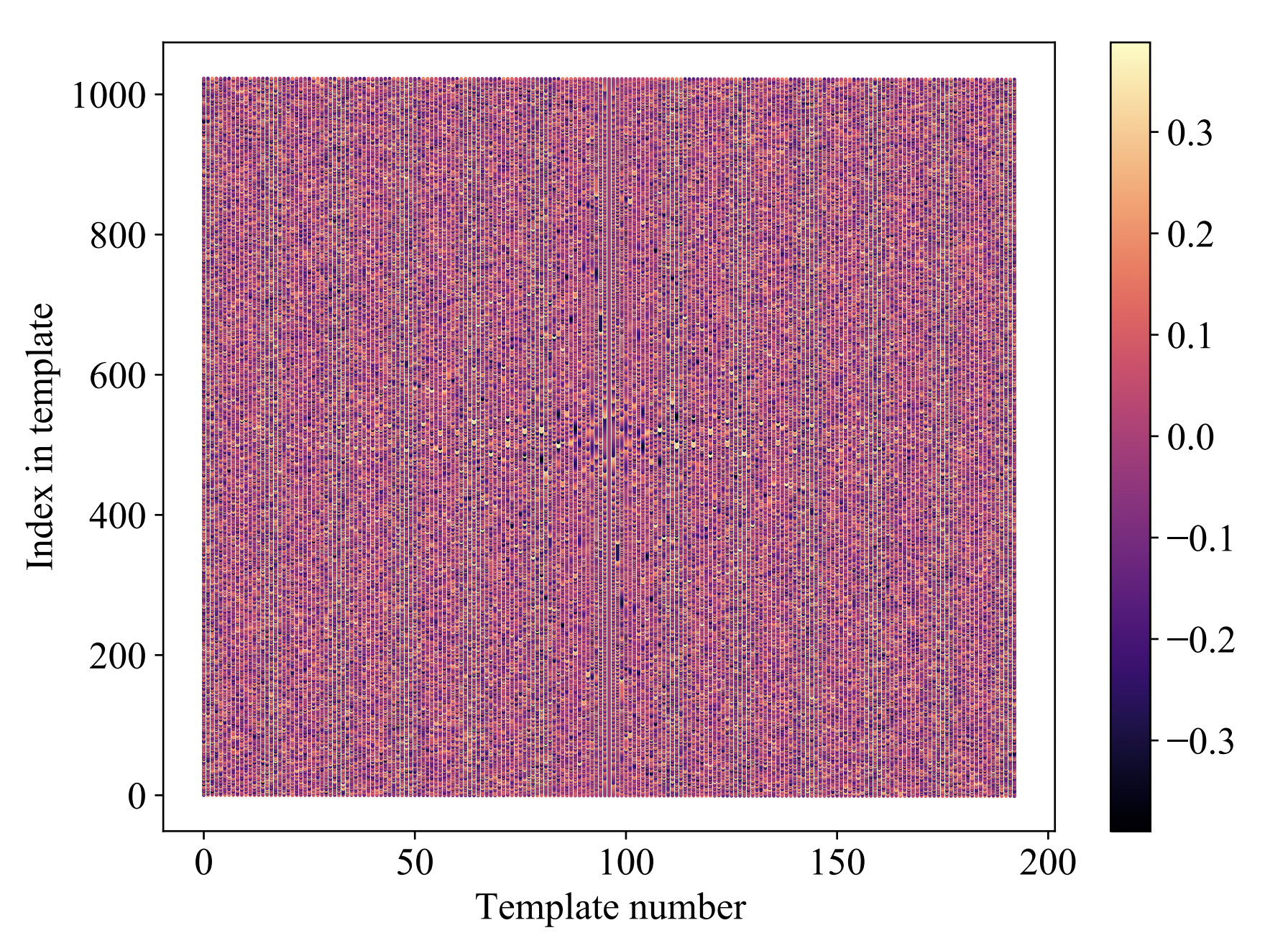}
\caption{Relative difference between imaginary components of acceleration templates, colour scale represents relative difference as in Equation \ref{eqn:perc_diff}.}
\label{imdiff_template}
\centering
\end{figure}

Where $b$ is the bfloat16 value and $s$ is the single-precision value. The result of applying this equation to the real and imaginary part of all templates is plotted in Figure \ref{rediff_template} and Figure \ref{imdiff_template}.

This metric will be reused in multiple experiments as it is important to have a comparison of bfloat16's ability to recreate higher precision results that is invariant to the size of the numbers being dealt with.

\begin{table}[htb]
\centering
\begin{tabular}{ |c||c|c|c|c| } 
\hline
Component & Mean & St. Dev & Max & Min\\
\hline \hline
Imaginary & -0.001569\% & 0.1537\% & 0.3883\% & -0.3894\% \\
\hline
Real & -0.001093\% & 0.1532\% & 0.3878\% & -0.3898\% \\
\hline
\end{tabular}
\caption{\label{tab:template_stats}Statistics of the difference between single-precision and bfloat16 after template conversion, aggregated from data used to generate Figure \ref{rediff_template} and Figure \ref{imdiff_template}}
\end{table}

The aggregated statistics of the relative differences are shown in Table \ref{tab:template_stats}, it is clear that at this early stage in the pipeline, only a very small amount of numerical error has been introduced by reducing the precision of the acceleration templates from single-precision to bfloat16.

\section{Methodology}
\subsection{Generating synthetic data}

SIGPROC \cite{2011ascl.soft07016L} was used to generate synthetic pulsar data. The input parameters to SIGPROC's \code{fake} determine the physical properties of the binary system being simulated.

To test and compare the varying precisions of AstroAccelerate across a representative range of potential pulsars, the \code{fake} input parameters are sampled from a log-uniform distribution with upper and lower bounds (Table \ref{tab:binary_fake}), so as to get an even representation of potential numerical scales, without introducing bias towards any particular range. The bounds are picked to represent a wide range of potentially physically viable pulsars, although some combinations of parameters are at the extrema of our ranges and are included to investigate the limits of the numerical processing.

Sampling a single point in this space is computationally costly, synthetic \code{fake} file generation followed by running the generated data through AstroAccelerate takes roughly 2 minutes. Therefore, it is desirable to reduce the dimensionality of the input parameter space wherever it is physically justifiable, to reduce the `curse of dimensionality'.

\begin{table}[htb]
\centering
\begin{tabular}{ |c|c|  }
 \hline
 Parameter Name& Range\\
 \hline \hline
 period&[1.25, 1000] \\
 width&[4, 50]\\
 snrpeak&[0.0125, 0.125]\\
 dm&[5.. 500]\\
 nbits&8\\
 nchans&1024\\
 tsamp&128\\
 tobs&600\\
 fch1&1550\\
 foff&0.292968752\\
 additional flags&-binary\\
 bper&[1.5, 336]\\
 bphase&0.2\\
 bpmass&[1.0, 1.5]\\
 bcmass&[0.1, 5.0]\\
 \hline
\end{tabular}
\caption{\label{tab:binary_fake}Configuration arguments for SIGPROC \code{fake} command to generate a binary pulsar signal. [a, b] represents the closed interval inclusive of the bounds, while [a.. b] represents the integer interval inclusive of the bounds}
\end{table}

\subsection{Selecting an optimal bphase value}

The \code{bphase} parameter to \code{fake} is a number in the range 0 to 1 which determines the starting phase of the binary orbit in the range 0\degree to 360\degree. To reduce the dimensionality of the parameter space, it is necessary to determine roughly which value of \code{bphase} leads to the largest acceleration value of the detected peaks in the output space, to ensure each test is of a pulsar signature with at least some acceleration. Conceptually, the largest observed Doppler shift is when the emitter is travelling directly towards or directly away from the observer.

\begin{figure}[htb]
\includegraphics[scale=0.22]{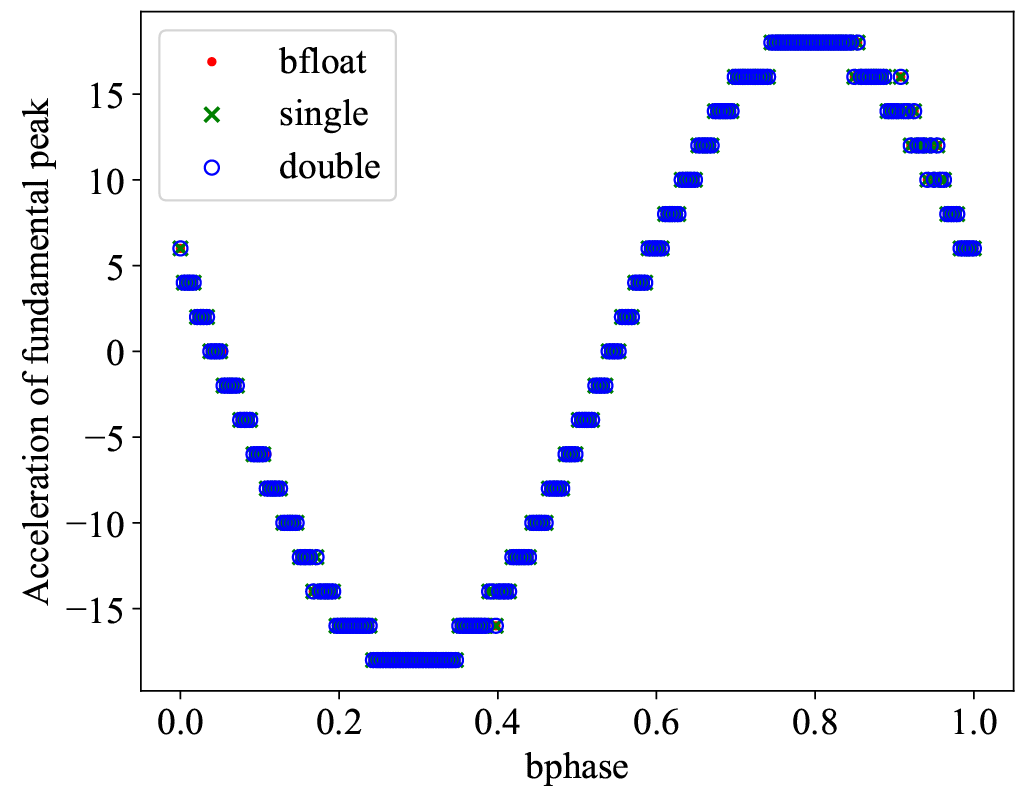}
\caption{Fundamental peak acceleration (`z' bin) against \code{bphase} in bfloat16, single and double-precision (overlayed)}
\label{accvsbphase}
\centering
\end{figure}

The simple experiment chosen is to hold all parameters fixed at typical values, and allow \code{bphase} to sweep over its entire range. Then, by plotting the output frequency-acceleration plane, the acceleration value of the highest peak can be captured, and plotted, Figure \ref{accvsbphase}.

Through the results of this sweep, the static value selected for \code{bphase} is 0.2, this approximately centres the observation window over the period of orbit with maximum linear acceleration towards the observer, as shown in Figure \ref{accvsbphase}. This is an approximation as the \code{bper} value (which sets the orbital period in hours) is a variable input parameter to \code{fake}, and it is the combination of \code{bphase} and \code{bper} that sets the midpoint of the observation window. 

\subsection{Peakfinding methodology}

An accelerated pulsar signal can be observed in the $f-\dot{f}$ plane as a series of peaks at integer multiples of $f$ and $\dot{f}$. Quantisation error in both the frequency and acceleration axes of the search lead to the observed peaks drifting into adjacent bins. To account for this, a harmonic sum algorithm is used to gather the power from all harmonics into a single detection. However, in this work we are interested in the location of each harmonic peak individually, so a harmonic sum is not used. 

A more general and naive peakfinding and harmonic separation approach is taken to enable us to look at each harmonic separately. Since the period of the synthetic pulsar is known a priori, being an input parameter to \code{fake}, it is possible to preemptively split the $f-\dot{f}$ output plane with boundaries (Table \ref{tab:harm_bounds}) derived from the fundamental frequency $f$, which is calculated as $f = \frac{1}{period}$Hz.

\begin{table}[htb]
\centering
\begin{tabular}{ |c|c|  }
 \hline
Band& Range /Hz\\
 \hline \hline
 Fundamental&[$0, 1.5 \times f$] \\
 First harmonic&[$1.5 \times f , 2.5 \times f$]\\
 Second harmonic&[$2.5 \times f, 3.5 \times f$]\\
 Nth harmonic&[$(N+0.5) \times f, (N+1.5) \times f$]\\

 \hline
\end{tabular}
\caption{\label{tab:harm_bounds}Harmonic band search bounds. [a, b] represents the closed interval inclusive of the upper bound, and lower in the case of the fundamental.}
\end{table}

Once the $f-\dot{f}$ plane has been split into bands based on the calculated fundamental frequency, the $n^{th}$ harmonic is taken to be the highest point in the $n^{th}$ band. In the case of low SNR pulsar signals, peaks and their harmonics are likely to be at similar powers to the background noise. This leads to a failure mode where the peak selection method picks a random point of noise as the harmonic in that band. As the noise is randomly distributed, the highest point in the bfloat16 plane and the single-precision plane are not guaranteed to be in the same position, leading to some extreme observed values of bin drift. This failure mode is discussed and quantified in section \ref{sec:bindrift}.

In the majority of cases where the peak height is not at the noise floor, measuring the ``bin drift" of a peak in a given band between precisions allows us to quantify the combined extent of the spectral leakage, accumulated numerical error and scalloping loss, and its effect on our ability to resolve a given pulsar from a certain noisy background.

The following section compares the ability of our FDAS implementation (in various precisions) to detect the presence of accelerated pulsars in synthetic datasets.

\section{Results}
In this section we present the results from analysing output $f-\dot{f}$ planes with randomly generated inputs, to understand the conditions under which bfloat16 is a good (or bad) approximation to the higher precision implementations. 

\subsection{Noise properties in varying precisions}

\begin{figure*}[htb]
  \makebox[\textwidth][c]{\includegraphics[width=\textwidth]{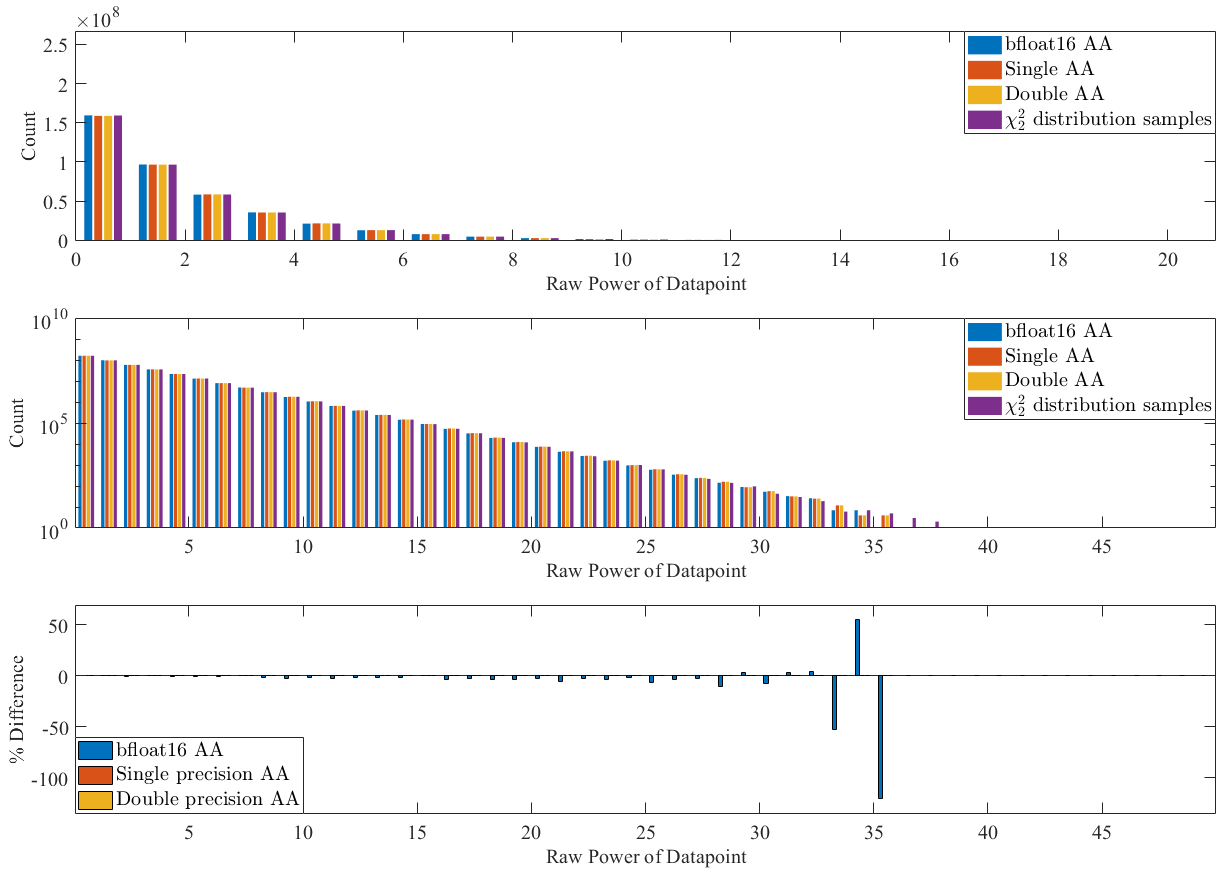}}%
  \caption{Raw noise power histogram with ${\chi}^2_2$ comparison, linear scale (top). Raw noise power histogram with ${\chi}^2_2$ comparison, log scale (middle). Percent difference between bfloat16 vs. single-precision bin count and double-precision vs single-precision bin count (bottom). In all data series, N = 404,750,143 samples.}
  \label{noise_hist}
\end{figure*}

To understand if a reduction in precision has any statistically significant effect on the background noise distribution, synthetic data without a pulsar signal is analysed in this section.

The histogram in Figure \ref{noise_hist} (top) represents the distribution of output power across an entire $f-\dot{f}$ plane for a filterbank file generated using a SIGPROC \code{snrpeak} parameter of 0. The \code{snrpeak} parameter can be thought of as proportional to the brightness of the signal. When \code{snrpeak} is 0, there is no signal in the file being generated. Between precisions, there is no significant skewing, and only very slight perturbations on the count in each bin (compared to the value in single-precision) which can be observed upon careful examination of Figure \ref{noise_hist} (bottom).

It is interesting to note, that however slight the perturbations are on the count in each bin, the overall trend is that there are marginally fewer large values in the bfloat16 noise than the single-precision noise. 

However, we conclude that reducing or increasing precision has no significant effect on the probability density function (PDF) that the noise is drawn from.

Figure \ref{noise_hist} shows that the noise distribution follows a ${\chi}^2_2$ distribution, this is expected as the input data without signal is well approximated by white noise.

\subsection{Synthetic dataset results - Power}

In this section we discuss the effect of changing precision on the ability of AstroAccelerate's implementation of FDAS to resolve the SNR of a particular peak, where a pulsar signal is known to be present in the dataset. SNR is defined as the raw power (height) of the peak on the $f-\dot{f}$ plane divided by the standard deviation of the entire plane.

\begin{center}
\begin{figure*}[htb]
\includegraphics[width=\textwidth]{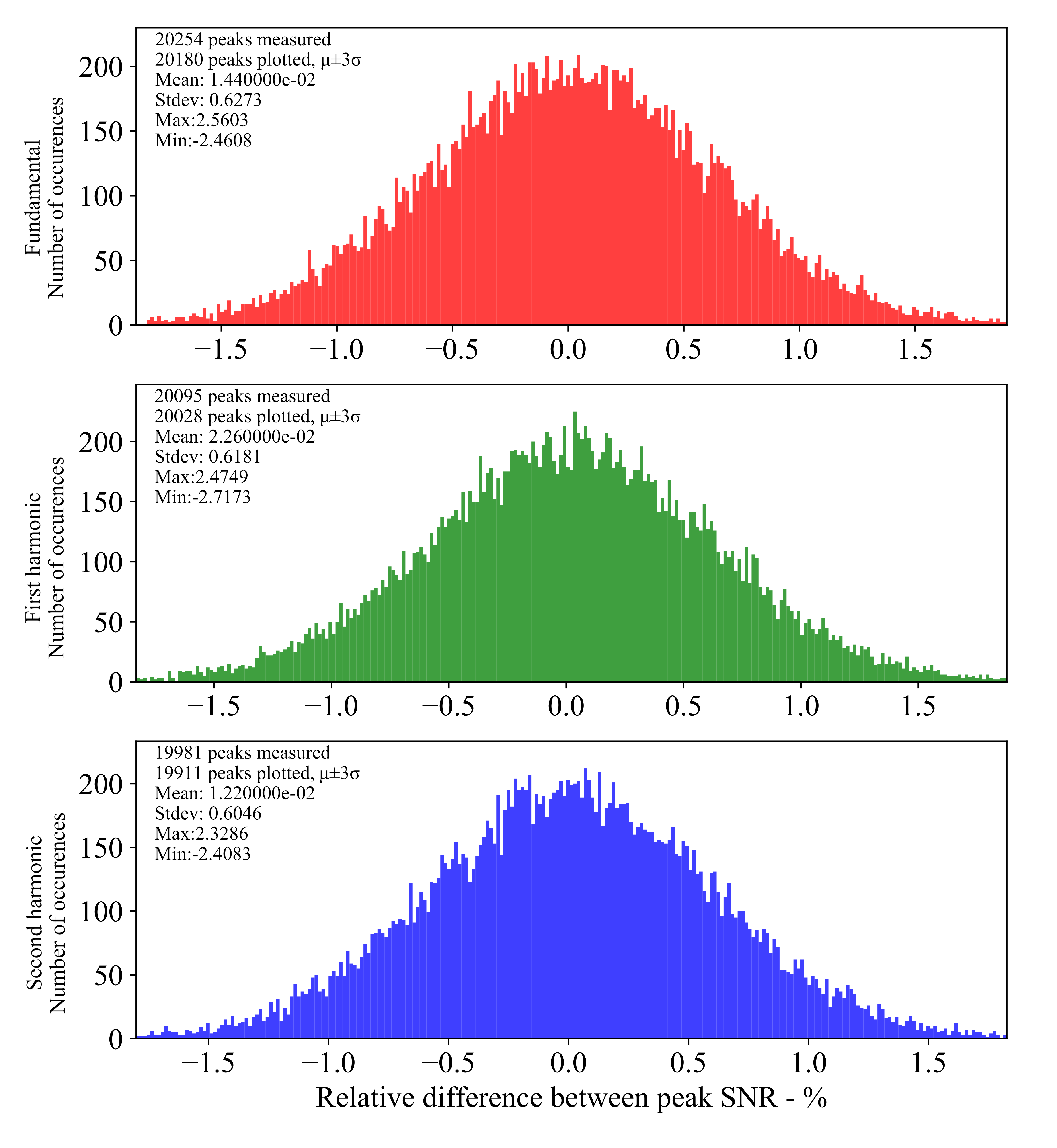}
\caption{Graphs depicting the spread in peak power recreation between bfloat16 vs. single-precision. }
\label{snr_peaks}
\end{figure*}
\end{center}

Figure \ref{snr_peaks} shows the relative difference in SNR between candidate peaks in bfloat16 and single precision. To calculate this we have taken the power of the highest peak in a given harmonic band in bfloat16, also taken the highest peak in the same harmonic band in the single precision output plane and compared their SNR via equation \eqref{eqn:perc_diff}. This process is repeated across many samples spanning the input parameter space (Table \ref{tab:binary_fake}) to quantify the limitations of the reduced precision mode.

The first three harmonics are analysed separately in each row of Figure \ref{snr_peaks}. There are three different precisions under investigation (bfloat16, single-precision, double-precision). The comparison in Figure \ref{snr_peaks} is made between bfloat16 and single-precision. The analysis of the equivalent single-precision vs double-precision data showed that for $98.8\%$ of the peaks there was no difference in height, and the maximum observed difference was $0.0098\%$. This is as we would expect, as rounding a result to single-precision eliminates the majority of the benefit of a preceding double-precision calculation, unless extreme amounts of error were accumulating in single-precision, for example if the algorithm was numerically unstable.

With a sufficiently large dataset, the distribution of peak power differences when comparing bfloat16 and single-precision output follows a normal distribution, centred on zero. In the first three harmonics, it is seen that the standard deviation is consistently $\approx 0.6\%$.

\subsection{Synthetic dataset results - Bin drift}
\label{sec:bindrift}

A reduced precision detection pipeline must not only be able to reproduce the detection SNR to within an acceptable error, but also localise the detection in the $f-\dot{f}$ plane correctly.

As such, it is important to understand whether reducing precision to bfloat16 also affects the location of a pulsar signature peak and its harmonics on the acceleration ($\dot{f}$) or frequency axis. 

The results of this section aim to quantify the impact of reducing precision in FDAS from single-precision to bfloat16 when measuring the frequency and acceleration ($\dot{f}$) properties of binary pulsar signatures.

\begin{center}
\begin{figure*}[htb]
\includegraphics[width=\textwidth]{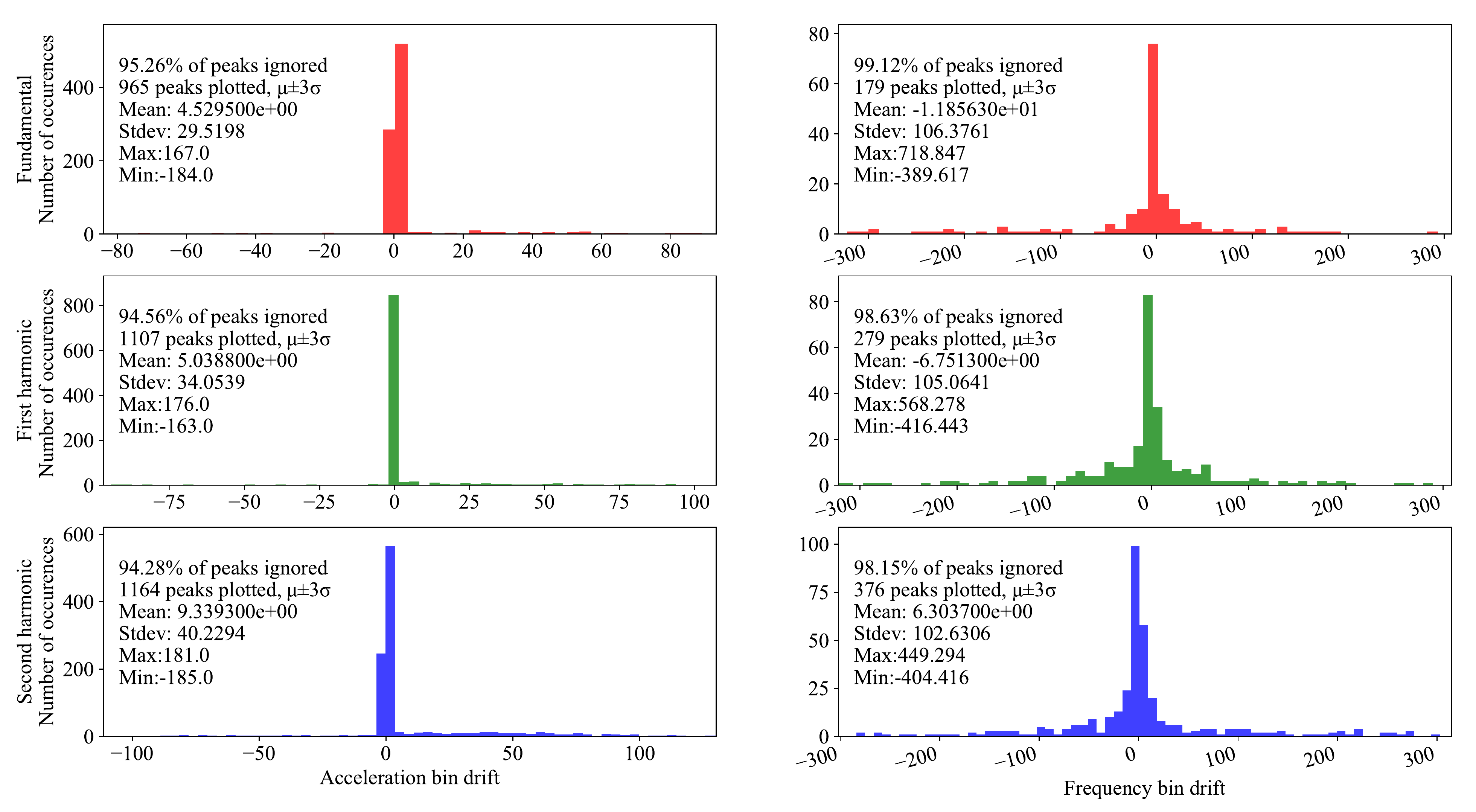}
\caption{Graphs depicting acceleration (left) and frequency (right) bin drift between bfloat16 and single-precision results.}
\label{freq_accn_b2s}
\end{figure*}
\end{center}

The bin drifts on the frequency and acceleration axes between bfloat16 and single-precision are shown in Figure \ref{freq_accn_b2s}. Across all tests performed there is no measurable bin drift between the single and double-precision versions of FDAS.

When comparing single-precision and bfloat16, the majority ($>94\%$) of peaks do not show any bin drift, this is shown by the large fraction of peaks that are ignored in Figure \ref{freq_accn_b2s}. Although in cases where there is bin drift, it is usually limited to adjacent bins, i.e. the peak in the bfloat16 band is ± 1 bins from the peak in single-precision. This is shown by the tall peak at the centre of the histograms. 

\subsubsection{Extreme bin drifts}

As previously stated, there are some cases where the bin drift is extreme, and looks to be far more than could be attributed to just accumulated numerical error. This is due to the limitations of the chosen peakfinding method mis-classifying noise as a peak in the response in files where the signal of the pulsar is extremely weak, below the noise floor.

\begin{figure*}[htb]
\includegraphics[width=\textwidth]{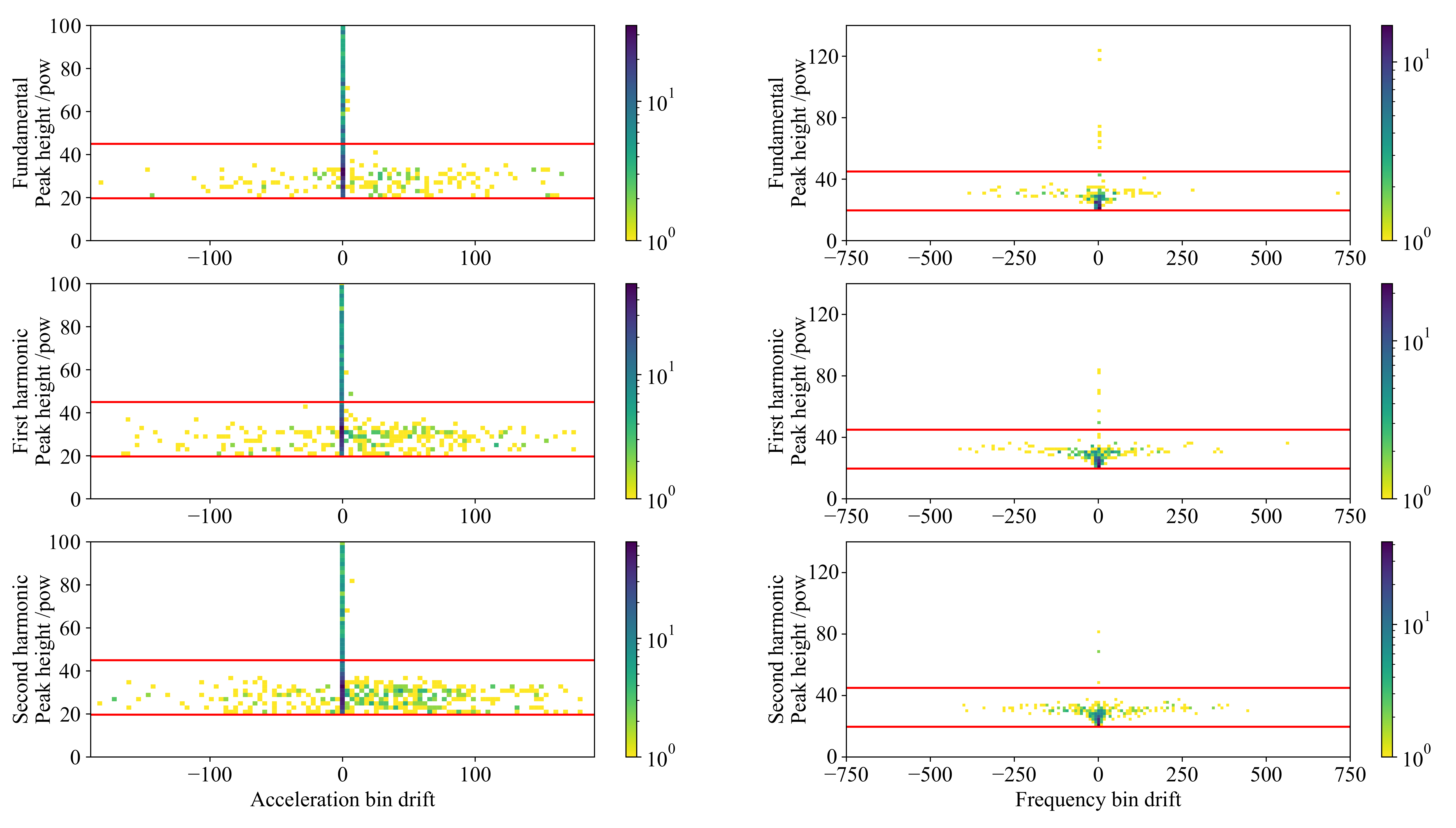}
\caption{2D Histograms depicting the spread of acceleration (left) and frequency (right) bin drift with peak power, colour scale represents absolute count. Upper and lower red bars represent measured maximum and minimum values of the largest noise value in noise-only data samples.}
\label{bin_drift_hist}
\centering
\end{figure*}

Figure \ref{bin_drift_hist} is important in understanding this particular limitation of our chosen analysis technique. Ignoring the red lines, this is a 2D histogram plot of the same data as in Figure \ref{freq_accn_b2s}. The y-axis represents the peak powers, so larger peaks are at the top of the histogram, and now the colour intensity represents the number of occurrences in a given bin drift/peak power bin. Again, points with zero bin drift are ignored, leaving a vertical strip of data at a bin drift of ±1, which is more noticeable in the acceleration plots. However in both the acceleration and frequency plots, there is a band of very large bin drifts at lower values of peak power, around the noise floor. The explanation of this effect follows, assisted by the red lines.

In cases where a peak lies in or below the canopy of the noise, the large shifts occur under the following conditions:

\begin{enumerate}
\item In each harmonic band under investigation, the two tallest peaks are very similar heights (in both bfloat16 and single-precision), within the range shown in Figure \ref{snr_peaks}.
\item AND, when changing between precisions, the peak that was the higher of the two in bfloat16 becomes the lower of the two peaks in single-precision, they ``flip".

\end{enumerate}

This is why it is only a small subset of many cases within the ``noise canopy" where the effect is seen. 

If both of the two highest peaks are due to noise, they could be located anywhere randomly in the harmonic band. This means that when the peakfinding methodology selects the ``harmonic" from the harmonic band it is effectively just picking two random locations, which can lead to a very large measured bin drift.

The proportion of cases which we attribute to this explanation, account for much less than 1\% of the dataset, and are extremely low SNR peaks.

We have designed and carried out an experiment to measure the location of the relevant noise canopy. The experimental method is as follows:

\begin{enumerate}
    \item Generate a selection of noise-only output $f-\dot{f}$ planes (without any pulsar signature present) by setting SIGPROC's \code{snrpeak} to 0, and process each one through the AstroAccelerate FDAS pipeline in single-precision mode.
    \item For each extreme bin drift event (magnitude $>$1) seen in Figure \ref{bin_drift_hist}, note the frequency (calculated as $f = \frac{1}{period}$Hz) of the pulsar that was in the synthetic data file that led to the extreme bin drift event. This frequency will have determined the width of the noise band the peak selection algorithm was picking from as in Table \ref{tab:harm_bounds}.
    \item For each extreme bin drift event, take an unused band of noise from the noise-only $f-\dot{f}$ planes with a frequency width of $1.5 \times f$, and record the highest point.
    \item Record the maximum and minimum of all of the ``highest points", these are the y-coordinates of the bounds (horizontal red lines) in Figure \ref{bin_drift_hist}.
\end{enumerate}

The bounds on Figure \ref{bin_drift_hist} demonstrate that all the observed extreme bin drift events occur in the region (the ``noise canopy") where we would expect to see the randomly located noise peaks. In this region a pulsar detection is already ambiguous, so we do not consider the performance of FDAS as being degraded by these edge cases.

\subsubsection{Acceleration Skewing}
A subtle skewing within the wide horizontal band in Figure \ref{bin_drift_hist} can also be observed, with values on the right hand side of the x-axis occurring slightly more frequently.

This skewing can be thought of as representing that when two peaks were confused between bfloat16 and single-precision, the bfloat16 peak had a more positive acceleration than the single-precision peak. While curious, this skewing does not affect our interpretation of the performance of bfloat16 FDAS for detecting pulsars.

Through constructing flat convolution templates (for example making Figure \ref{single_acc_template} a flat plane) in both single-precision and bfloat16, and reperforming the experiment with a comparison template where the imaginary component is perturbed by a single bfloat16 numbers width, we have established that there is no asymmetric accumulation of error between +ve and -ve acceleration templates (where the simulated pulsar is travelling towards or away from the observer) during the convolution process.

This leads us to believe the effect arises as a subtle result of the average complex magnitude of the input templates, in which there is an extremely slight (at the sixth significant figure) but measurable increase in average power across the templates corresponding to positive acceleration values compared to the negative templates. We believe this is why the skew is rare and may only appear in cases where we have established that the potentially confused peaks are already extremely close, so only a tiny increase in average power would be enough to cause a trend towards one side.

\subsection{Performance}

\begin{figure}[htb]
\includegraphics[scale=0.2]{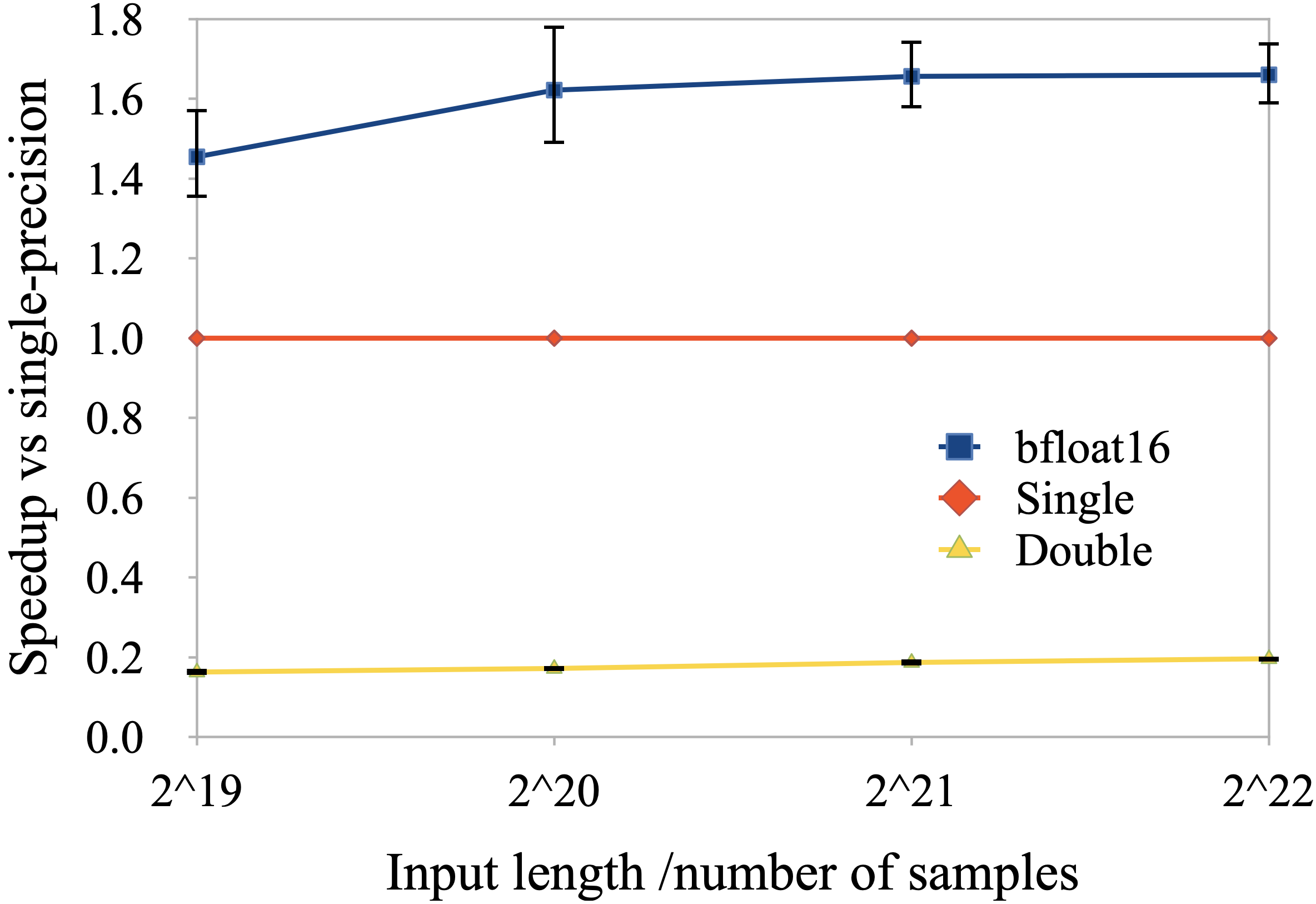}
\caption{GPU performance comparison with varying precision and input length (number of samples) in AstroAccelerate}
\label{performance}
\end{figure}

In Figure \ref{performance} we present the speedup of the GPU accelerated section of AstroAccelerate's implementation of FDAS with varying precision, this is all steps within the blue box in Figure \ref{fdas_flow}. This section represents a variable fraction of the overall execution time of a binary pulsar search, depending on the number of templates and number of dispersion measures (DMs) analysed. 

Measurements were taken with varying input data length (measured in number of samples). Each datapoint represents the average speedup of 256 runs, calculated by comparing the execution time of the modified precision version with the average execution time of the corresponding single-precision version. The error bars are the measured standard deviation of the speedup.

\subsection{Real data}
In Table \ref{tab:real_data_power} we present the ability of our bfloat16 version of AstroAccelerate to detect a pulsar in data from a real telescope, the GMRT. We used an observation of a 2.16 ms pulsar (J1544+4937) in a 2.8 hour compact orbit. The 550-750 MHz filterbank data was dedispersed at the nominal dispersion measure of 23.23 pc cm$^{-3}$ \cite{osti_22136604}, generating a time-series with a resolution of 81.92$\mu$s. An observation time of $345s$ was used to compare the bloat16 pipeline to the corresponding single and double-precision versions, these results are also compared to those from PRESTO. 

We believe the discrepancy in the power values between PRESTO and AA is caused by scalloping loss. Due to slight variations in the exact number of samples processed by PRESTO and AA (owing to limitations on input length to varying FFT algorithms), the frequency bin centres will be offset between the results.

Importantly, the reported acceleration and frequency values match for all the three choices of numerical precision of AstroAccelerate.

Linear regression on the first three peaks from the AstroAccelerate and PRESTO output leads to a measured acceleration of the pulsar of 2.839 ms$^{-2}$ (AA) and 2.931 ms$^{-2}$ (PRESTO).

\section{Conclusions}
This work has demonstrated that it is possible to reduce the precision of the convolution section of the FDAS binary pulsar detection pipeline.

We have compared three different precisions, bfloat16, single-precision floating point (IEEE-754) and double-precision floating point (IEEE-754) across a wide range of binary system parameters, including real data of a millisecond pulsar in a compact orbit. We have found that the benefit of using double-precision is very limited, and usually does not result in a measurable difference when the results are rounded to single-precision (as may be required for later processing). Given the increased memory bandwidth requirements of doubling the precision, these results certainly do not justify increasing precision in the FDAS pipeline of AA.

Contrastingly, there is a small but measurable difference between the output of the bfloat16 implementation of FDAS and the single-precision implementation. However in all cases the peak power is within a few percent of the single-precision result, with no strong bias to be either higher or lower. This should enable users of GPU accelerated FDAS to increase their pulsar parameter search space on a given set of hardware when performing a real time search. Or for a given search, reduce their hardware requirement, when compared to using the single-precision implementation. 

\section*{Acknowledgements}
This work has received support from STFC Grant (ST/T000570/1). The authors would like to acknowledge the use of the University of Oxford Advanced Research Computing (ARC) facility in carrying out this work \cite{richards_2015_22558}. The authors would also like to express their gratitude to the Research Centre for Theoretical Physics and Astrophysics, Institute of Physics, Silesian University in Opava for institutional support.

The National Radio Astronomy Observatory is a facility of the National Science Foundation operated under cooperative agreement by Associated Universities, Inc. SMR is a CIFAR Fellow and is supported by the NSF Physics Frontiers Center awards 1430284 and 2020265.

\begin{table*}[htb]
\centering
\begin{tabular}{ |p{1.7cm}||p{3.7cm}||p{3.9cm}||p{3.6cm}||p{3.8cm}|  }
 \hline
Harmonic&PRESTO Power&AA Power - bfloat16&AA Power - Single&AA Power - Double \\
 \hline
 \hline
 Fundamental&   188.07&    132&    132.277&    132.227\\
 1&             71.96&    103&     103.258&     103.259\\
 2&             47.4545&    48.5&     48.777&     48.777\\
 3&             18.7041&    18&   17.773&      17.773\\
 \hline
 \hline 
 \hline
Harmonic&PRESTO `z' Bin&AA `z' Bin - bfloat16&AA `z' Bin - Single&AA `z' Bin - Double\\
 \hline
 \hline
 Fundamental&   0.75&   0.5&0.5&0.5\\
 1&             1.5&    1.25&1.25&1.25\\
 2&             1.5&    1.5&1.5&1.5\\
 3&             2&      2&2&2\\
 
\hline
Acceleration&2.931ms$^{-2}$&2.839ms$^{-2}$&2.839ms$^{-2}$&2.839ms$^{-2}$\\

 \hline
 \hline
 \hline
Harmonic&PRESTO Frequency&AA Frequency - bfloat16&AA Frequency - Single&AA Frequency - Double\\
 \hline
 \hline
 Fundamental&463.098Hz&463.097Hz&463.097Hz&463.097Hz\\
 1&926.196Hz&926.192Hz&926.192Hz&926.192Hz\\
 2&1389.29Hz&1389.29Hz&1389.29Hz&1389.29Hz\\
 3&1852.39Hz&1852.39Hz&1852.39Hz&1852.39Hz\\
\hline
\end{tabular}
\caption{\label{tab:real_data_power}Comparison of J1544+4937 GMRT Data, processed with PRESTO and AstroAccelerate in varying numerical precisions. Power values taken from the $f-\dot{f}$ plane in each program.}
\end{table*}

\bibliography{main}
\bibliographystyle{aasjournal}

\end{document}